\documentclass[aps,prd,showpacs,showkeys]{revtex4}
\usepackage{amsmath}
\usepackage{graphicx}

\begin{document}
\title{Thermodynamics of Constant Curvature Black Holes Through Semi-Classical Tunneling}
\date{February 24, 2011}

\author{Alexandre Yale}
\email{ayale@perimeterinstitute.ca}
\affiliation{Perimeter Institute, 31 Caroline St. N., Waterloo, Ontario N2L 2Y5, Canada}

\begin{abstract}
We study the semi-classical tunneling of scalar and fermion fields from the horizon of a Constant Curvature Black Hole, which is locally AdS and whose five-dimensional analogue is dual to ${\cal N}=4$ Super Yang-Mills.  In particular, we highlight the strong reliance of the tunneling method for Hawking radiation on near-horizon symmetries, a fact often hidden behind the algorithmic procedure with which the tunneling approach tends to be used.  We ultimately calculate the emission rate of scalars and fermions, and hence the black hole's Hawking temperature.
\end{abstract}

\maketitle
\newcommand{\defeq}[2]{\begin{equation} \begin{split} \label{#1} #2 \end{split} \end{equation}}
\newcommand{\myeq}[1]{\begin{equation} \begin{split} #1 \end{split} \end{equation}}

\section{Introduction}
Studying the thermodynamic properties of black holes is often regarded as a fundamental step towards quantizing gravity.  Indeed, the realization that a black hole's entropy is related to its area is at the root of the gauge / gravity correspondence which forms the basis of much modern research.  Although most of the literature focuses on dualities between conformal field theories and Anti de Sitter space --- and to a lesser extent on extremal Kerr and similar black holes --- it is expected that a much wider range of spacetimes embody a holographic principle of a similar nature.  As we study more general classes of spacetimes, we might gain access to more general classes of field theories.  As such, it is imperative to develop tools to study the thermodynamics of spacetimes for which the standard methods, such as the Wick rotation, may not be appropriate.
\\\\
A particular approach to study Hawking radiation, called the tunneling method, has recently garnered a lot of attention.  Not only does it paint a very physical picture of Hawking radiation --- that of quantum fields tunneling through the event horizon --- but it is also local in nature.  This property makes the technique versatile enough to apply to a wide range of black holes, such as embedded black holes in de Sitter spacetimes.  As such, it has been used to investigate the thermodynamics of a wide variety of spacetimes, many of which previously difficult or impossible to study.  They include Kerr-Newman \cite{Jiang2006,Zhang2006}, Black Rings \cite{Zhao2006}, Taub-NUT \cite{Kerner2006}, AdS Black Holes \cite{Hemming2001}, BTZ \cite{Agheben2005,BTZ1,BTZ2,BTZ3}, Vaidya \cite{Vaidya}, dynamical black holes \cite{Cri2007}, Kerr-G\"{o}del \cite{Kerner2007}, and deSitter horizons \cite{dS1,dS2,dS3,dS4,dS5,dS6}, as well as generic weakly isolated horizons \cite{Wu2007}.
\\\\
Another particularly important advantage of the tunneling approach is its ability to consider quantum fields explicitly.  While it has so far only been used to study scalars --- including some coupled to a $U(1)$ gauge field \cite{Zhu2011} --- spin $1/2$ \cite{Kerner2008,Kerner2008b,Li2008,Zhang2006,Jiang2006} and higher-spin \cite{Yale2009} fermions, and bosons \cite{Yale2010b,Majhi3}, one can expect the method to eventually mature and be used to study, for example, fields whose equations of motion come from effective string theoretic actions.
\\\\
The tunneling method has a rich history.  Based on the earlier works of Volovik \cite{Volovik1,Volovik2,Volovik3}, who analyzed properties of superfluidic Helium-3 related to quantum tunneling, Kraus and Wilczek \cite{NullGeo1,NullGeo2,NullGeo3} studied Hawking radiation semiclassically by studying modes of the field near the event horizon.  This was later generalized as a tunneling phenomenon by Parikh and Wilczek \cite{PW}.  Padmanabhan and his collaborators finally developed a second flavour of this method \cite{PaddyTun1,PaddyTun2,PaddyTun3,PaddyTun4} --- this is the version we will use here --- which is now starting to become well understood.
\\\\
The idea behind the method can easily be illustrated using a free massless scalar field being emitted from a Schwarzschild black hole.  Writing the field $\phi = e^{\frac{i}{\hbar} I}$, where $I$ is its action, the field equations reduce to first order in $\hbar$ to the Hamilton-Jacobi equations: $\partial_\mu I \partial^\mu I = 0$.  In order to solve this equation, we will need to make use of the spacetime symmetries, namely that $\partial_t I$, $\partial_\theta I$ and $\partial_\phi I$ are constants of motion.  Hence, we may impose that the action obeys the ansatz $I = -Et + W(r) + J_1 \theta + J_2 \phi$, which forces the Hamilton-Jacobi equations to reduce to $W'(r) = \frac{\pm E}{1 - 2M/r}$.  Finally, this means that the black hole's temperature is given by $\beta = 8 \pi M$, since the field's emission rate is given by
\myeq{
\Gamma \propto e^{- \text{Im} I} \propto e^{-2 \text{Im} \oint W'(r) dr} \propto e^{-8 \pi M E}.
}
The Constant Curvature Black Hole \cite{Banados1998} can be thought of as a $3+1$ dimensional extension to the BTZ black hole, and was found by identifying points in anti-deSitter spacetime.  Using Schwarzschild-like coordinates \cite{Cai2002}, the line element can be written as
\defeq{eq:CCBH}{ds^2 = \frac{l^4 f^2(r)}{r_H^2} \left[ d \theta^2 - \sin^2 \theta (dt/l)^2 \right] + \frac{dr^2}{f^2(r)} + r^2 d \phi^2,}
where
\defeq{eq:f}{ f^2(r) = \frac{r^2 - r_H^2}{l^2};}
the horizon is located at $r=r_H$.  Because of its non-standard asymptotic limit, all conserved charges diverge.  To make sense of the global charges, Ba\~{n}ados initially studied the thermodynamics of the CCBH by embedding it in a Chern-Simons supergravity theory.  Later, using quasi-local thermodynamic variables to tame the divergences, Creighton and Mann \cite{Creighton1998} studied the thermodynamics in a General Relativistic setting, where they showed that the first law of thermodynamics still holds.
\\\\
We study this type of black hole for two reasons.  First, black holes with an unusual topology --- the CCBH is $R^3 \times S^1$ instead of the more usual $R^2 \times S^2$ --- might be dual to interesting field theories, and form prime candidates for spacetimes which should be fully understood.  In particular, Cai \cite{Cai2002} has shown that the five dimensional CCBH is dual to ${\cal N}=4$ Super-Yang-Mills. Second, this black hole will highlight a connection between the tunneling method and near-horizon symmetries which is often overlooked; indeed, it is often outright assumed, such as in Kerr \cite{Li2008,Kerner2008b}, Kerr-AdS \cite{Agheben2005} and Taub-NUT-AdS \cite{Kerner2006}, that the outgoing radiation is purely radial near the horizon: the quantum field tunnels out with constant $\theta=\theta_0$.  Yet this connection is crucial: these near-horizon symmetries are precisely those which might suggest the existence of a conformal field theory at the horizon.
\\\\
We will begin by calculating the Hawking temperature of the CCBH using the Wick rotation method, for comparison purposes.  Then, we will apply the tunneling method to the problem and show that one must account for the near-horizon symmetries of the black hole in order to guess the right form of the action's ansatz.  This will lead us to the black hole's temperature.

\section{Constant Curvature Black Hole}
\subsection{Wick rotation}
Because the CCBH is a fairly simple spacetime, its temperature can easily be calculated using a Wick rotation.  This will provide us with a reference against which to compare the temperature we will calculate using the tunneling method.  Using the CCBH metric in the form of equation $(\ref{eq:CCBH})$, we notice that we require $t/l$ to have a period of $2 \pi$ in order to avoid a conical singularity in the Euclideanized metric; this association means that the term in square brackets is simply the line element of a two-sphere.  Hence, the inverse temperature is $\beta = 2 \pi l$.
\\\\
Note that if we denote the timelike Killing vector by $t^a = \delta^a_t$, then $\nabla^a t^b \nabla_a t_b = \frac{2}{l^2 r_H^2} \left(r^2 \sin^2 \theta + r_H^2 \cos^2 \theta \right)$; this means that the surface gravity is $\kappa = 1/l$ and, therefore, $\beta = 2 \pi l$ once again.
\\\\
Because our spacetime is not asymptotically flat, we should for clarity redshift the inverse temperature by multiplying by $\sqrt{-g_{00}}$.  The temperature observed by an observer located at $(r,\theta)$ is finally given by
\defeq{WickTemp}{
T = \frac{r_H}{2 \pi l^2 f(r)\sin \theta} .
}

\subsection{Tunneling}
\begin{figure}[t]
	\centering
		\includegraphics[width=0.25\textwidth]{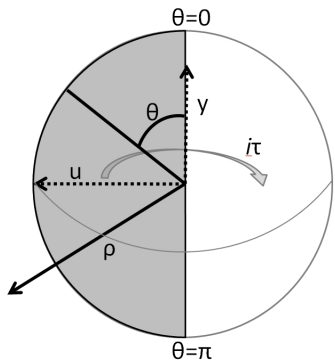}
		\caption{A constant time-slice of the near-horizon CCBH, with the angular coordinate $\phi$ suppressed.  The axisymmetric axis, formed by the $\theta=0$ and $\theta=\pi$ lines, represents the horizon.  The radiation emitted from this black hole should follow the coordinate $u = \rho \sin \theta$.}
	\label{fig:Figure}
\end{figure}

In contrast with the above calculations, the tunneling approach to Hawking radiation has a direct physical interpretation: that of a quantum field tunneling through an event horizon.  For completeness, we will simultaneously consider both the emission of a free scalar field $\phi$, obeying the Klein-Gordon equation, and the emission of a free fermion field $\Psi$.  In general, the equations of motion for this fermion should be given by the Rarita-Schwinger equations \cite{Rarita1941}.  However, it has been shown \cite{Yale2009} that both types of fields are emitted from black holes at the same rate, thus yielding the same Hawking temperature.  Therefore, we will assume that our fermions obey the Dirac equation, and we can write:
\begin{equation} \begin{split} \label{eom}
\left( \nabla_\mu \partial^\mu + m^2 \right) \phi &= 0 \\
\left(i \gamma^\mu D_\mu + m \right) \Psi &= 0 ,
\end{split} \end{equation}
where the covariant derivative $D_\mu$ acting on the fermion field is defined by
\begin{equation}	
	D_\mu = \partial_\mu - \frac{1}{8}\Gamma^{\alpha \phantom{\mu} \beta}_{\phantom{\alpha}\mu}[\gamma_\alpha,\gamma_\beta].
\end{equation}
The $\gamma^\mu$ matrices satisfy $\left\{ \gamma^\mu,\gamma^\nu \right\} = 2g^{\mu \nu}$.  We will write them in the form $\gamma^\mu = e_I^\mu \hat{\gamma}^I$, where we've introduced tetrads satisfying $e_I^\mu e_J^\nu \eta^{I J} = g^{\mu \nu}$ as well as the $\hat{\gamma}$ matrices which correspond to the flat-space Dirac matrices in the basis
\begin{equation} \begin{split}
\hat{\gamma}^t &= \left( \begin{array}{cc} 0 & 1 \\ -1 & 0 \end{array} \right)  \quad
\hat{\gamma}^r = \left( \begin{array}{cc} 0 & \sigma^3 \\ \sigma^3 & 0 \end{array} \right) \\
\hat{\gamma}^\theta &=\left( \begin{array}{cc} 0 & \sigma^1 \\ \sigma^1 & 0 \end{array} \right)  \quad
\hat{\gamma}^\phi = \left( \begin{array}{cc} 0 & \sigma^2 \\ \sigma^2 & 0 \end{array} \right) .
\end{split} \end{equation}
The $\sigma^i$ correspond to the standard Pauli matrices:
\begin{equation} \begin{split}
\sigma^1 = \left( \begin{array}{cc} 0&1\\1&0 \end{array} \right) \quad
\sigma^2 = \left( \begin{array}{cc} 0&-i\\i&0 \end{array} \right) \quad
\sigma^3 = \left( \begin{array}{cc} 1&0\\0&-1 \end{array} \right).
\end{split} \end{equation}
Using the WKB approximation, we impose that our fields be proportional to $e^{\frac{i}{\hbar}I}$, where $I$ is the action.  For the scalar field, this means that $\phi = C(x^\mu)e^{\frac{i}{\hbar}I_s}$.  For the fermion, we focus on the spin-up case (the spin-down case being effectively identical) and denote  the positive-spin eigenvector of $\sigma^3$ by $\xi$.  The field $\Psi$ then takes the form
\begin{equation} \begin{split}
\Psi = \left[ \begin{array}{c} \xi A(x^\mu) \\ \xi B(x^\mu) \end{array} \right] e^{\frac{i}{\hbar} I_{\uparrow}(x^\mu)}
= \left[  \begin{array}{ c}     A(x^\mu)\\     0\\     B(x^\mu)\\     0  \end{array} \right] e^{\frac{i}{\hbar} I_{\uparrow}(x^\mu)}.
\end{split} \end{equation}
Because the metric is invariant under translations over $t$ and $\phi$, we know that $\partial_t I \equiv -E$ and $\partial_\phi I \equiv J$ will be conserved quantities.  As such, our action will take the form
\begin{equation} \label{actionansatz}
	I = -Et + W(r,\theta) + J \phi.
\end{equation}
Then, inputting the fields $\phi$ and $\Psi$ into their respective equations of motion $(\ref{eom})$ and using equation $(\ref{actionansatz})$, we find for the scalar
\begin{equation} \label{scalareom}
\frac{ -r_H^2 E^2}{l^2 f^2(r) \sin^2(\theta) } + f^2(r) W_r^2(r,\theta) + \frac{r_H^2}{l^4 f^2(r)} W_\theta^2(r,\theta) + \frac{J^2}{r^2} + m^2 = 0,
\end{equation}
and for the fermion:
\begin{equation} \begin{split}
 \label{fermionEquations}
	-B \left( \frac{-E r_H}{l \sin(\theta) \sqrt{f^2(r)}} + \sqrt{f^2(r)} W_r(r,\theta) \right) + mA = 0 \\
	-B \left( \frac{r_H}{l^2 \sqrt{f^2(r)}} W_\theta (r,\theta) +  \frac{iJ}{r} \right) = 0 \\
	-A \left( \frac{ E r_H}{l \sin(\theta) \sqrt{f^2(r)}} + \sqrt{f^2(r)} W_r(r,\theta) \right) + mB = 0 \\
	-A \left( \frac{r_H}{l^2 \sqrt{f^2(r)}} W_\theta (r,\theta) +  \frac{iJ}{r} \right) = 0,
\end{split} \end{equation}
where we approximated to leading order in $\hbar$ and subsequently set $\hbar$ to unity.  To this order, we effectively have $\nabla_\mu \approx D_\mu \approx \partial_\mu$.  Similarly, derivatives of $A$ and $B$ are also of higher order in $\hbar$, making $A$ and $B$ effectively constant.  Although at first glance this approximation appears to remove a lot of interesting physics, it has recently been shown that the terms of higher order in $\hbar$ will generate no contribution to the Hawking temperature \cite{Post2,Yale2010b,Mitra2}.  As such, we are well justified in truncating our expressions to leading order in $\hbar$.
\\\\
Near the horizon, $f^2(r)$ goes to zero, such that the mass term does not contribute.  This allows us to rewrite the fermion equations as
\begin{equation} \begin{split} \label{fermionsCCBH}
	W_r(r,\theta) = \frac{ \pm E r_H}{l \sin(\theta) f^2(r)} \\
	W_\theta (r,\theta) = \frac{ i J l^2 \sqrt{f^2(r)}}{r r_H},
\end{split} \end{equation}
while the equations for the scalar are still given by equation $(\ref{scalareom})$; equations of the same form as $(\ref{fermionsCCBH})$ are also found in multiple other spacetimes \cite{Li2008,Kerner2008b,Kerner2006}.  There, the next step is to assume that the Hawking radiation is purely radial: that is, $\theta = \theta_0$ during the tunneling process.  Unfortunately, this assumption, which is often made without justification, does not work in the case of the CCBH.  Indeed, it implies that $W_\theta$ will yield no contribution to $\text{Im} I$, and integrating $W_r(r)$ across the horizon, for both the scalar and the fermion, will yield
\begin{equation}
	\text{Im} W \propto  \frac{ \pm \pi l E}{2 \sin( \theta_0) }.
\end{equation}
This expression cannot be correct, as it would ultimately imply that the Hawking temperature depends on the emission coordinate $\theta_0$ (which describes where on the horizon the field has been emitted from), which contradicts our previous calculations from equation $(\ref{WickTemp})$ (where the Hawking temperature only depended on the observer's position).
\\\\
In order to proceed forward, we need to understand how to decompose this $W(r,\theta)$ into two functions, each depending on only one variable.  In that spirit, we will take a short detour to discuss near-horizon symmetries before coming back to these field equations later.  As a first step, we take the near horizon limit and rewrite the line element, initially given by equation $(\ref{eq:CCBH})$, into a more explicit form:
\defeq{eq:nearHorizon}{ ds^2= \rho^2(d \theta^2 - \sin^2 \theta d \tau^2) + d \rho^2 + dz^2,}
where we've introduced the coordinates
\defeq{defs}{
\tau = \frac{E}{l} \quad
\rho = \frac{f(r)l^2}{r_H} \quad
z = r_H \phi.
}
Note that the horizon is now located at $\rho=0$.  This line element corresponds to an observer with constant acceleration in Minkowski space.  Indeed, if we introduce the coordinates $u =\rho \sin \theta$ and $y = \rho \cos \theta$, then the near-horizon line element becomes
\defeq{eq:Standard_Rindler}{ds^2 = -u^2 d \tau^2 + d u^2 + d y^2 + d z^2.}
The near-horizon symmetries are now immediately obvious, and it is clear that $\xi_1^\mu = \delta_\tau^\mu$, $\xi_2^\mu = \delta_y^\mu$, and $\xi_3^\mu = \delta_z^\mu$ are Killing vectors.  We can therefore define three conserved quantities $E = -\partial_t I$, $J_1 = \partial_y I$, and $J_2 = \partial_z I$ which correspond to the energy of the emitted field, as well as its momentum along the $y$ and $z$ directions.  Therefore, near the horizon, the action will have the form
\defeq{good_ansatz}{
I &= -E t + W(u) + J_1 y + J_2 z \\
&= -E t + W \left( \frac{f(r)l^2}{r_H} \sin \theta \right) + J_1 \frac{f(r)l^2}{r_H} \cos \theta + J_2 r_h \phi,
}
where, on the second line, we've written the near-horizon coordinates $(u,y,z)$ in terms of the general coordinates $(r,\theta,\phi)$ from equation $(\ref{eq:CCBH})$.  It is important to note that this construction is only valid in the near-horizon limit.  We are, however, justified in taking such a limit because of the nature of the tunneling method.  Indeed, as we will see in the next section, all contributions to the Hawking temperature come from discontinuities located directly at the horizon.  This is precisely the power of the tunneling method: because the near-horizon geometry is so much simpler than the bulk geometry, additional symmetries arise which greatly simplify the field equations that we are trying to solve.
\\\\
The field should be understood not as being emitted radially along the $r$ direction while keeping $\theta$ constant, but as being emitted along the $u=\rho \sin \theta$ direction keeping $y=\rho \cos \theta$ constant.  As can be seen in Figure \ref{fig:Figure}, one can think of the CCBH as a sphere with an event horizon at $\rho=0$ with additional coordinate singularities at $\theta=0$ and $\theta=\pi$.  While $\rho$ parametrizes the $r=r_H$ horizon, $\rho \sin \theta$ corresponds to the distance from the axisymmetry axis which itself forms a coordinate singularity.  It is from this axisymmetric line that the particles will be emitted.  A discussion of the CCBH instanton can also be found in \cite{Creighton1998}.
\\\\
Knowing that our action takes the simple form from equation $(\ref{good_ansatz})$, we can return to our field equations.  The scalar field still obeys the same equation $(\ref{scalareom})$ as before:
\begin{equation} \label{HJ2}
\frac{ -r_H^2 E^2}{l^2 f^2(r) \sin^2(\theta) } + f^2(r) I_r^2(r,\theta) + \frac{r_H^2}{l^4 f^2(r)} I_\theta^2(r,\theta) + \frac{I_\phi^2}{r^2} + m^2 = 0.
\end{equation}
For the fermion, on the other hand, we should rotate the field because we've modified the definition of our radial coordinate.  However, we can save ourselves from having to do this by being clever.  Indeed, the fermion field will now look like
\begin{equation}
\Psi = \left[  \begin{array}{ c}     E(x^\mu)\\     F(x^\mu)\\      G(x^\mu)\\      H(x^\mu)  \end{array} \right] e^{\frac{i}{\hbar} I_{\uparrow}(x^\mu)};
\end{equation}
inputting this into the Dirac equation will yield a system of equation similar to equation $(\ref{fermionEquations})$.  However, notice that because we are working in a first-order approximation, all these equations are linear in $\Psi$ (Interestingly, this trick works even without truncating to first order in $\hbar$ \cite{Yale2010b}).  Writing the equations in matrix form $M \Psi = 0$, therefore, we demand that the determinant of $M$ vanishes; if it did not, then the only solution would be $\Psi=0$.  This gives the equation
\begin{equation}
0 = \text{Det} M = \frac{ -r_H^2 E^2}{l^2 f^2(r) \sin^2(\theta) } + f^2(r) I_r^2(r,\theta) + \frac{r_H^2}{l^4 f^2(r)} I_\theta^2(r,\theta) + \frac{I_\phi^2}{r^2} + m^2,
\end{equation}
which is the same equation as that for the scalar field $(\ref{HJ2})$.  Using the ansatz for the action $(\ref{good_ansatz})$ into these equations of motion $(\ref{HJ2})$, we get
\begin{equation} 0 = \frac{ -r_H^2 E^2}{l^2 f^2(r) \sin^2\theta} + \left[W' \left( \frac{f(r)l^2}{r_H} \sin \theta \right)\right]^2 + J_1^2 + \frac{J_2^2}{r^2} + m^2, \end{equation}
where $W'$ denotes the derivative of $W$ with respect to its entire argument.  Because we are working near the horizon, $f^2(r)$ goes to zero such that the first two terms diverge while the last three are finite.  Hence,
\begin{equation} \label{Wprime} W'_{\pm} \left( \frac{f(r)l^2}{r_H} \sin \theta \right) = \frac{ \pm r_H E}{l f(r) \sin\theta}, \end{equation}
where the $+$ and $-$ correspond to outgoing and infalling radiation, respectively.  More details on this correspondence can be found in Section 3 of \cite{Yale2010b}.

\subsection{Calculating the temperature}
Given the ``radial'' derivative of the action, given by equation $(\ref{Wprime})$, it is straightforward to calculate the imaginary part of the action, which is related to the Hawking temperature through the tunneling probability $\Gamma \propto e^{- \text{Im} I} = e^{-E/T}$.  While there are a number of techniques which have been introduced for this step \cite{Mitra2006,Akhmedov2006,Akhmedov2007,Akhmedov2008,Stotyn2009,Chowdhury2008a,Chowdhury2008b,Pilling2008}, we will here follow the ideas recently summarized in \cite{Gill2010}.  In particular, it is found that there are two separate contributions to the imaginary part of the action: one coming from the discontinuity of the time coordinate $\Delta t$, and another coming from a pole, at the horizon, which arises when integrating the radial part of the action.  The temperature is ultimately given by the expression
\begin{equation}
\label{SingletonTemp} T = \frac{E}{\text{Im} \left(\int W'_+ - \int W'_- - 2 E \Delta t \right)},
\end{equation}
where the $\int W'_{\pm}$ must each be integrated along an infinitesimal quarter-circle path around the pole at the horizon \cite{Gill2010}, such that the result of the integration will be $\frac{i \pi}{2}$ times the residue at the pole.  Hence:
\begin{equation} \begin{split} \label{W_Temp}
\int W'_+ d \left( \frac{f(r)l^2}{r_H} \sin \theta \right) &= \frac{i \pi}{2} l E\\
\int W'_- d \left( \frac{f(r)l^2}{r_H} \sin \theta \right) &= \frac{-i \pi}{2} l E.
\end{split} \end{equation}
Next, we need to calculate the jump in the time coordinate across the horizon $\Delta t$.  Notice that the line element defined by $(\ref{eq:nearHorizon})$ is that of an accelerated observer in Minkowski space following the path
\defeq{RindlerPath}{
T &= \rho \sin \theta \sinh \tau \\
X &= \rho \sin \theta \cosh \tau \\
Y &= \rho \cos \theta \\
Z &= r_H \phi
}
outside the horizon, whereas inside the horizon, we need to interchange $X \leftrightarrow T$ to account for the timelike coordinate becoming spacelike as the horizon is crossed.  Thus, when we cross the horizon, we must have $\tau \rightarrow \tau - \frac{i \pi}{2}$ in order to get $\sinh \tau \leftrightarrow \cosh \tau$.  The time discontinuity is therefore $\Delta t = l \Delta \tau = \frac{-i \pi}{2} l$.
\\\\
We can combine this result with equations $(\ref{SingletonTemp})$ and $(\ref{W_Temp})$ to find the Hawking temperature:
\begin{equation}  \label{finalTemp}
T = \frac{E}{\pi E l + \pi E l} =\frac{1}{2 \pi l}.
\end{equation}
Redshifting to account for the observer's position, this becomes
\begin{equation} \label{finalT}
T = \frac{r_H}{2 \pi l^2 f(r)\sin \theta} ,
\end{equation}
which is the same temperature as the one found using the Wick Rotation method $(\ref{WickTemp})$.

\section{Conclusions}
We have calculated the first-order Hawking temperature for fermions and scalars emitted from the Constant Curvature Black Hole by considering the process as a semi-classical tunneling phenomenon across the event horizon.  This spacetime emphasizes a few interesting aspects of the tunneling method which are not obvious in simpler black hole spacetimes, such as the necessity to carefully choose an ansatz for the action based on the symmetries of the near-horizon geometry.  This suggests a link between this technique and gauge / gravity dualities in black hole spacetimes, which also rely on these symmetries, that should be the focus of further studies.  Moreover, our calculation serves as a practical demonstration of a new idea \cite{Gill2010} for accounting for the factor of 2 problem --- by which the tunneling method yielded differing emission probabilities depending on the type of coordinates used --- by considering the discontinuity of the time coordinate across the event horizon.

\acknowledgments{
This work was supported by the Natural Sciences and Engineering Research Council of Canada.  The author would like to thank Robert Mann for pointing out the CCBH spacetime, as well as Ross Diener and Nima Doroud for insightful comments on the manuscript.
}


\begin{thebibliography}{99}
\bibitem{Zhang2006} J Zhang and Z Zhao,	Phys. Lett. B 638, 2-3 (2006) 110 [gr-qc/0512153]
\bibitem{Jiang2006} Q Jiang, S Wu and X Cai, Phys. Rev. D 73 (2006) 064003 [hep-th/0512351]
\bibitem{Zhao2006} L Zhao, Commun. Theor. Phys. 47 (2007) 835 [hep-th/0602065] 
\bibitem{Kerner2006} R Kerner and R B Mann, Phys. Rev. D 73 (2006) 104010 [gr-qc/0603019] 
\bibitem{Hemming2001} S Hemming and E Keski-Vakkuri,  Phys. Rev. D 64 (2001) 044006 [gr-qc/0005115]
\bibitem{Agheben2005} M Agheben, M Nadalini, L Vanzo and S Zerbini, JHEP 0505 (2005) 014 [hep-th/0503081] 
\bibitem{BTZ1} S Wu and Q Jiang, JHEP 0603 (2006) 079 [hep-th/0602033]; 
\bibitem{BTZ2}R Li and J Ren, Phys. Lett. B 661, 5 (2008) 370 [arXiv:0802.3954]; 
\bibitem{BTZ3}W Liu, Phys. Lett. B 634, 5-6 (2006) 541 [gr-qc/0512099]
\bibitem{Vaidya} R Jun, Z Jing-Yi and Z Zheng, Chin. Phys. Lett. 23 (2006) 2019 [gr-qc/0606066] 
\bibitem{Cri2007} R Di Criscienzo et al, Phys. Lett. B 657, 1-3 (2007) 107 [arXiv:0707.4425]
\bibitem{Kerner2007} R Kerner and R B  Mann, Phys. Rev. D 75 (2007) 084022 [hep-th/0701107]
\bibitem{dS1} M K Parikh, Phys. Lett. B 546 (2002) 189 [hep-th/0204107];
\bibitem{dS2}A J M Medved, Phys. Rev. D 66 (2002) 124009 [hep-th/0207247]; 
\bibitem{dS3}S Shankaranarayanan, Phys. Rev. D 67 (2003) 084026 [gr-qc/0301090]; 
\bibitem{dS4}S Wu and Q Jiang (2006)	[hep-th/0603082]; 
\bibitem{dS5}Y Sekiwa (2008)	[arXiv:0802.3266]; 
\bibitem{dS6}D Chen, Q Jiang and X Zu (2008)	[arXiv:0804.0131] 
\bibitem{Wu2007} X Wu and S Gao, Phys. Rev. D75	(2007) 044027 [gr-qc/0702033] 
\bibitem{Zhu2011} T Zhu (2011) [arXiv:1101.4466]
\bibitem{Kerner2008} R Kerner and R B Mann, Class. Quantum Grav. 25 (2008) 095014 [arXiv:0710.0612] 
\bibitem{Li2008} R Li, J R Ren and S W Wei, Class. Quantum Grav. 25 (2008) 125016 [arXiv:0803.1410] 
\bibitem{Kerner2008b} R Kerner and R B Mann, Phys. Lett. B665 (2008) 277 [arXiv:0803.2246] 
\bibitem{Yale2009} A Yale and R B Mann, Phys. Lett. B 673, 2 (2009) 10 [arXiv:0808.2820] 
\bibitem{Yale2010b} A Yale, Phys. Lett. B (In Press) [arXiv:1012.3165]
\bibitem{Majhi3} B R Majhi, S Samanta, Ann. Phys. 325	(2010) 2410 [arXiv:0901.2258] 
\bibitem{Volovik1} G E Volovik, JETP Lett. 69 (1999) [gr-qc/9901077] 
\bibitem{Volovik2} G E Volovik, ``Topological Defects and the Non-Equilibrium Dynamics of Symmetry Breaking Phase Transitions'', Y.M. Bunkov and H. Godfrin, Kluwer Academic Publishers (2000) [cond-mat/9902171]
\bibitem{Volovik3} G E Volovik, ``Exotic properties of superfluid 3He'', World Scientific, Singapore (1992)
\bibitem{NullGeo1} P Kraus and F Wilczek, Mod. Phys. Lett. A9, 40 (1994) 3713 [gr-qc/9406042];
\bibitem{NullGeo2} P Kraus and F Wilczek, Nucl. Phys. B433 (1995) 403 [gr-qc/9408003];
\bibitem{NullGeo3} P Kraus and F Wilczek, Nucl. Phys. B437 (1995) 231 [hep-th/9411219] 
\bibitem{PW} M K Parikh and F Wilczek, Phys. Rev. Lett. 85 (2000) 5042 [hep-th/9907001] 
\bibitem{PaddyTun1} K Srinivasan and T Padmanabhan, Phys. Rev. D 60 (1999) 24007 [gr-qc-9812028];
\bibitem{PaddyTun2} S Shankaranarayanan, K Srinivasan and T Padmanabhan, Mod.Phys.Lett. A16 (2001) 571 [gr-qc/0007022];
\bibitem{PaddyTun3} S Shankaranarayanan, T Padmanabhan and K Srinivasan, Class.Quant.Grav. 19 (2002) 2671 [gr-qc/0010042v4];
\bibitem{PaddyTun4} T Padmanabhan, 	Mod. Phys. Letts. A 19 (2004) 2637 [gr-qc/0405072]
\bibitem{Banados1998} M Banados, Phys. Rev. D 57, 2 (1998) 1068 [gr-qc/9703040]
\bibitem{Cai2002} R G Cai, Phys. Lett. B544, 1 (2002) 176 [hep-th/0206223]
\bibitem{Creighton1998} J D E Creighton and R B Mann, Phys. Rev. D 58 (1998) 024013 [gr-qc/9710042]
\bibitem{Rarita1941} W Rarita and J Schwinger, Phys. Rev. 60, 61 (1941) 
\bibitem{Post2} A Yale (2011) [arXiv:1102.5102]
\bibitem{Mitra2} B Chatterjee and P Mitra, Phys. Lett. B675 (2009) 240 [arXiv:0902.0230]
\bibitem{Akhmedov2008} 	E Akhmedova, T Pilling, A de Gill and D Singleton, Phys. Lett. B666 (2008) 269 [arXiv:0804.2289]
\bibitem{Mitra2006} P Mitra, Phys. Lett. B648 (2007)240	[hep-th/0611265] 
\bibitem{Stotyn2009} S Stotyn, K Schleich and D Witt, Class. Quant. Grav. 26 (2009) 065010 [arXiv:0809.5093] 
\bibitem{Akhmedov2006} E T Akhmedov, V Akhmedova and D Singleton, Phys.Lett. B642 (2006) 124 [hep-th/0608098]
\bibitem{Akhmedov2007} E T Akhmedov, V Akhmedova, T Pilling and D Singleton, Int.J.Mod.Phys.A22:1705 (2007) [hep-th/0605137]
\bibitem{Chowdhury2008a} B D Chowdhury, 	Pramana70 (2008) 593
\bibitem{Chowdhury2008b} B D Chowdhury, 	Pramana70 (2008) 3 [hep-th/0605197]
\bibitem{Pilling2008} T Pilling, Phys.Lett.B660 (2008) 402 [arXiv:0709.1624]
\bibitem{Gill2010} A de Gill, D Singleton, V Akhmedova and T Pilling, Amer J Phys 78, 7 (2010) 685 [arXiv:1001.4833] 

\end{thebibliography}
\end{document}